# Mechanical cues for totipotency and the preneural state: embryo and cancer expanding the frontiers of developmental physics

Jaime Cofre*

Laboratory of Molecular Embryology and Cancer, Federal University of Santa Catarina, room 313b, Florianópolis, SC, 88040-900, Brazil

* Corresponding author. Laboratory of Molecular Embryology and Cancer, Universidade Federal de Santa Catarina, sala 313b, Florianópolis, SC, 88040-900, Brazil

E-mail address: jaime.cofre@ufsc.br

**Abstract**

In this article, I advance the idea that physics plays a central role in cell differentiation and makes fundamental contributions to morphogenesis, revealing the totipotent nature of the zygote. Totipotency is a persistent mechanical memory that preserves the biomechanical records of animal morphogenesis. I examine the mechanical and biophysical pathways underlying cell differentiation in embryonic development and cancer, treating them as closely related biological and mechanical processes. Drawing inspiration from evolutionary history, I also propose a biophysical mechanism for the emergence of the animal nervous system. By linking physical principles to cellular differentiation, this review positions mechanobiology as a pillar of innovation with high-impact clinical implications for diseases such as cancer.

**Keywords:** cell differentiation, default state of neural induction, developmental physics, embryology, totipotency.

**1. Historical foundations of chemistry as the preferred explanatory model in embryology**

Classical embryological models of pattern formation disregard mechanical morphogenesis (Turing 1952), focusing instead on chemical morphogenesis (Gierer and Meinhardt 1972). From my perspective, a central limitation of developmental biology has been the failure to seek the fundamental mechanisms of self-organization and asymmetry in the first animal embryo or even in basal animals. This approach has led to the consolidation of classical embryological models that remain largely detached from an evolutionary understanding of early animal phylogeny (Wolpert 1969, 1971). Nevertheless, I identified points of convergence between chemical models and mechanical morphogenesis. Alan Turing believed that morphogens require the specification of only two states, a novel state and a default state (Turing 1952). Chemical models consider short-range activators and long-range inhibitors (Turing 1952; Gierer and Meinhardt 1972). Mechanical phenomena generate long-range tensile forces (Serra-Picamal et al. 2012) and compositional effects on the extracellular matrix (ECM) that alter the diffusion coefficients of bone morphogenetic protein (BMP) (Sedlmeier and Sleeman 2017), resembling short-range activators. Furthermore, small random molecular fluctuations may locally increase activator concentration (Gierer and Meinhardt 1972). Mechanical waves and tissue-level textural fluctuations fit perfectly within the expectations of the chemical model of morphogens (Turing 1952). Thus, I propose that a physical approach to the embryonic development of basal animals will, in the near future, provide significant advances in developmental biology.

I would also like to make an initial reflection on Turing's original formulation of the chemical bases of morphogenesis, in which he stated that he intended to focus on "cases where the mechanical aspect can be ignored and the chemical aspect is the most significant" (Turing 1952). It is very difficult to find embryological (Zhang et al. 2011; Tlili et al. 2019) or cancer-related (Hirway et al. 2021) phenomena where mechanical aspects can be truly ignored. For instance, the striking parallels between human gastrulation and cancer are well recognized, both of which begin with the disruption of the basal lamina. At the time of Turing's work, the focus on genes as the explanatory factor of morphogenesis was justified by the challenges and limitations of the period. Genes were understood as carriers of biological information. Today, however, we have a growing theoretical and experimental foundation of the physics of embryonic development, revealing the basis of totipotency (Hove et al. 2003; Nowlan et al. 2008; Adamo et al. 2009; North et al. 2009; Dado et al. 2012; Tajik et al. 2016; Vining and Mooney 2017; Engstrom et al. 2018; Nelson 2022; Wang et al. 2023; Nelson et al. 2024; Prado-Mantilla et al. 2025). This body of knowledge has helped us to understand that physical forces themselves are universal and do not evolve. The biological structures that generate or transmit those forces may change, but the underlying physical mechanisms remain unchanged. Chordin and collagen IV regulating BMPs have the same effect on nervous system induction. Physical phenomena are modular, context-specific in their interactions within the embryo, environmentally modifiable, and absolutely determinant in the evolution of animal forms and cell differentiation processes (Cofre 2025).

The intricate relationship between oncology and embryology is indisputable (Whitney 1901; Krebs 1947; Coggin and Anderson 1974; Pierce 1983), and cellular differentiation acts as the main point of convergence between the two fields (Markert 1968; Pierce and Johnson 1971). Consistent with this idea, Garry B. Pierce developed a concept of cancer grounded in embryological and oncological principles (Pierce and Wallace 1971). Pierce's concept of cancer anticipated a predictive value for embryogenesis (Pierce and Speers 1988). That is, his original idea was the starting point for recognizing that cancer and embryo are part of an inseparable process (Pierce and Wallace 1971; Pierce 1983; Arechaga 2003), like two sides of the same coin (Cofre and Saalfeld 2023), as well as for understanding the embryo as a differentiated benign tumor (Cofre 2024). Therefore, it is not possible to discuss embryonic cell differentiation without validating it within the field of oncology, and vice versa. Surprisingly, both cancer (Paszek and Weaver 2004; Paszek et al. 2005; DuFort et al. 2011) and embryogenesis (gastrulation in particular) (Brunet et al. 2013; Mitrossilis et al. 2017; Pukhlyakova et al. 2018; Bailles et al. 2019) can be triggered by mechanical forces and physical phenomena. Thus, the discussion on cancer presented in this article is essential for establishing a counterpoint between these closely related processes. Additionally, it is necessary to clarify the implications of developmental physics (a field not yet fully consolidated) for embryogenesis and to highlight recent efforts to translate mechanobiological insights into therapeutic strategies and clinical trials, an approach that Pierce himself referred to as differentiation therapy (Pierce and Speers 1988).

In what follows, we will broadly discuss how the physics of the morphogenetic construction of the embryo is responsible for revealing the totipotency of the zygote.

## 2. Differentiation: cues for biophysical and mechanical processes in embryology

Thinking about cell differentiation from a physics perspective allows us to understand that morphogenetic changes occur simultaneously with cell differentiation during embryogenesis. Therefore, the type of organization and embryonic construction has a direct impact on differentiation. What evidence is there that the transmission of an extrinsic force can influence gene expression? A good example of this influence is found in the linker of nucleoskeleton and cytoskeleton (LINC) complex, which comprises nesprins and Sad1/UNC-84 (SUN) proteins. LINC connects the nucleus to the cytoskeleton and plays a key role in the regulation of mechanical force transduction, which has been shown to have significant effects on global transcriptional regulation (Alam et al. 2016). Human adipose-derived stem cells cultured in hydrogels with stiffness gradients ranging from 0.5 to 8.2 kPa/mm showed stiffness-dependent expression of YAP, lamin A/C, lamin B, MRTF-A, and MRTF-B (Hadden et al. 2017). This effect was even more pronounced in human mesenchymal stem cells grown on rigid matrices, which exhibited extensive chromatin remodeling, including increased histone acetyltransferase activity, reduced histone deacetylase activity, and preserved integrity of the LINC complex (Killaars et al. 2019, 2020). On the other hand, mechanical tension can induce transcriptional repression and drive a switch in histone modification patterns in constitutive heterochromatin, from dimethylation of lysine 9 in histone H3 (H3K9me2/3) to trimethylation of lysine 27 in histone H3 (H3K27me3). This process is mediated by polycomb repressive complex 2 (PRC2), emerin, and non-muscle myosin IIA (Le et al. 2016).

Nuclear lamins provide another example of how external signals are transmitted to the nucleus, regulating transcription and functioning as mechanotransducers. The nuclear lamina of mammals is composed of intermediate filament proteins, including A/C- and B-type lamins, which are the primary determinants of nuclear rigidity. These lamins interact both directly and indirectly, via chromatin-binding adapter proteins, with specific genomic regions known as lamina-associated domains. Through these interactions, lamins play critical roles in chromatin organization, gene expression regulation, and the modulation of mechanosensitive transcription factors, such as NF-κB, a key mediator of immune function, inflammation, and cancer (Lammerding et al. 2004; Lund et al. 2013; Gruenbaum and Foisner 2015; Gesson et al. 2016).

A decade ago, Tajik et al. (2016) traced the full chain of mechanotransduction events from the cell surface to the nucleus, demonstrating that local tensions applied to integrins propagate through the strained actin cytoskeleton to the LINC complex, and from there through lamina–chromatin interactions to stretch chromatin and upregulate transcription. In fact, the mechanical forces transmitted from cell adhesions to the nucleus result in chromatin "stretching" within seconds of force application (Sun et al. 2020). Lamina-associated polypeptide 2β (LAP2β) has been implicated in transmitting forces from the nuclear lamina to chromatin, representing a mechanical force-induced mechanism of gene regulation (Sun et al. 2023). Further evidence comes from research by Emma Carley and colleagues, who demonstrated that the LINC complex regulates epidermal differentiation in vitro and in vivo. The authors showed that integrin-generated tension is communicated through the LINC complex to the nuclear lamina, preserving the progenitor state of basal keratinocytes. This finding suggests a direct mechanism by which mechanical signals can regulate cell differentiation (Carley et al. 2021).

Numerous studies with stem cells have demonstrated that the nanostructured morphology of the cell surface is a key factor promoting neurogenesis. For example, highly spiked surface moieties are particularly effective in inducing neuronal differentiation (Poudineh et al. 2018). Substrate geometry and dimensions (topography) also appear to influence whether stem cells differentiate into neurons or glial cells (Ankam et al. 2013). Other studies have shown that cell shape, cytoskeletal tension, and RhoA signaling proteins are critical determinants of human mesenchymal stem cell (hMSC) differentiation into adipocytes or osteoblasts (McBeath et al. 2004). Remarkably, remodeling of the actin cytoskeleton can transduce mechanical stimuli that promote muscle cell differentiation through RhoA/ROCK signaling (Huang et al. 2012). Furthermore, hMSCs subjected to functional blockade of vinculin fail to express myoblast determination protein 1 (MyoD), whereas expression of Runt-related transcription factor 2 (RUNX2), a key regulator of osteoblast differentiation, is stimulated, indicating that adhesion proteins sensitive to mechanical forces can directly influence the fate of stem cells (Holle et al. 2013). What is clear from all these findings, which strongly suggest a role of mechanics in cell differentiation, is the need to explore these mechanical phenomena in an embryonic context in vivo.

Understanding the factors that direct germ layer organization during embryogenesis is one of the main goals of developmental biology. In a study using atomic force microscopy to quantify the adhesive and mechanical properties of the ectoderm, mesoderm, and endoderm in zebrafish progenitor cells, Michael Krieg showed that the differential tension of the actomyosin-dependent cellular cortex (regulated by Nodal/TGF-β signaling) constitutes a key factor in germ layer organization during gastrulation (Krieg et al. 2008). More recently, it was found that the initial length of somites in zebrafish depends on a single mechanical property of the paraxial mesoderm: its surface tension (Naganathan et al. 2022). The formation of body segments in vertebrate embryos has long been attributed to the spatiotemporal pattern of chemical molecular signals (Cooke and Zeeman 1976; Palmeirim et al. 1997; Kawakami et al. 2005; Vermot and Pourquié 2005). However, we are now beginning to consolidate the understanding that somite length is defined by tissue mechanics (Packard and Jacobson 1979; Bard 1988; Grima and Schnell 2007; Nelemans et al. 2020).

The pioneering studies of Adam Engler and Dennis Discher were ahead of their time. They demonstrated that mesenchymal stem cells, influenced by ECM elasticity and MyoII, commit to neural, muscular, or osteogenic lineages. Soft matrices that mimic the brain promote neurogenesis, slightly stiffer matrices that mimic muscle promote myogenesis, and very stiff matrices induce osteogenesis (Engler et al. 2006). Recent findings have revealed that substrate stiffness can directly influence the three-dimensional organization of the genome, thereby regulating specific gene expression programs. Rigid substrates, similar to bone tissue, induce chromatin condensation and gene expression profiles associated with osteogenic differentiation (Na et al. 2025). This mechanotransduction pathway, in which mechanical signals are transmitted to the nucleus to induce nuclear deformations and chromatin remodeling, provides a direct link between the physical properties of the ECM and the activation of specific transcriptional programs. In my view, phenomena causing mechanical tension ultimately alter cell and tissue elasticity and vice versa, resulting in the activation of mechanotransduction mechanisms that are directly involved in cell differentiation.

Thus, the mechanical environment seems to be a determinant factor in regulating cell differentiation within the embryo. In other words, cells can sense differences in embryonic microenvironments and respond by differentiating into distinct cell types and tissue organizations (Wang et al. 1993; Gumbiner 1996, 2005). An important advance in our understanding of the mechanics of cell differentiation stems from studies assessing the development of the mouse epidermis (Prado-Mantilla et al. 2025). The formation of the skin barrier requires rapid proliferation along with differentiation and stratification of the embryonic epidermis. In the cited study, the authors demonstrated that an increase in intracellular actomyosin A contractility is sufficient to trigger terminal differentiation of embryonic skin, providing in vivo evidence of the interdependence between cell mechanics and differentiation.

A significant portion of the mechanical pathways guiding cell differentiation has been revealed through studies of tissue contacts and how they transmit and transform mechanical tension during embryogenesis. For example, studies on tension-induced mechanotransduction in *Caenorhabditis elegans* demonstrated that muscle cells influence epithelial morphogenesis (Zhang et al. 2011). Contacts between muscle and epithelium were proposed to occur via hemidesmosomes, which serve not only as anchoring structures but also as mechanosensors that respond to tension. These contacts initiate signaling cascades involving Rac GTPase, p21-activated kinase (PAK-1), and the adaptor proteins Git-1 and PIX-1 (Zhang et al. 2011). Mechanical forces generated by muscle contraction (muscle strength) are also crucial for bone growth in mouse embryos (Palacios et al. 1992) and help initiate and propagate ossification in long bones of avian embryos (Nowlan et al. 2008). In basal animals, particularly ctenophores, which have been placed at the root of animal phylogeny (Ryan et al. 2013; Moroz et al. 2014; Arcila et al. 2017; Shen et al. 2017; Whelan et al. 2017), there is no direct evidence of the influence of muscle cells on comb row formation. However, a connection between gonadal tissues and comb row formation has been suggested. Gametogenic tissues are located in the eight meridional canals underlying comb rows, or in their homologs or derivatives (Pianka 1974). The endodermal origin of the gonads is well established, and the organization of gonadal tissue allows distinguishing the different types of ctenophores (Pianka 1974).

Tissue contact and interaction, as seen in embryonic induction, represent some of the most fundamental processes in modern embryology (Spemann and Mangold 1924; Gurdon 1987; Jessell and Melton 1992; Slack 1993; Hemmati-Brivanlou and Melton 1997; Harland 2000). In basal animals, several cues for differentiation can be observed in the contacts between tissues during morphogenesis, following classical morphogen induction mechanisms. Although studies on the biophysics and mechanobiology of ctenophores are still lacking, understanding differentiation in this group is crucial within the context of this article, as I will later propose a model for nervous system formation based on basal animals.

It is well known that, in ctenophore embryos, micromere cells that move by epiboly differentiate into epidermal cells, whereas macromere cells subject to stress give rise to endodermal cells. The formation of comb rows requires inductive signaling between the $m_1$ and $e_1$ lineages (Martindale and Henry 1997). Classical inductive signals from E and M macromeres were described by Henry and Martindale (2001). Oral micromeres, on the other hand, migrate and undergo mesenchymal-to-epithelial transition (MET) to form muscle cells. If oral micromeres come into contact with the aboral epidermal surface, they differentiate into lithocytes of the apical organ (Martindale and Henry 1999). These classical inductive signals are well documented in ctenophores (Farfaglio 1963; Martindale 1986; Martindale and Henry 1996, 1997), as are the "embryonic fields or equivalence groups that are controlled by cell–cell interactions" (Henry and Martindale 2004).

In ctenophores, gastrulation generates the ectoderm through epiboly of aboral micromeres, whereas the endoderm forms by invagination (embolism) of macromeres that move inward, carrying with them a subset of micromeres into the interior of the embryo. These internalized micromeres, referred to as gastrular (Farfaglio 1963) or oral (Martindale and Henry 2015) micromeres, originate from macromeres (Pianka 1974). They appear on the oral side and are subsequently internalized during macromere invagination. Studies in *Mnemiopsis leidyi* indicated that most oral micromeres serve as progenitors of apical organ lithocytes (Freeman and Reynolds 1973) and give rise to mesodermal musculature (Martindale and Henry 1999). However, their potential contribution to endodermal derivatives cannot be excluded due to technical limitations (Martindale and Henry 1999). The muscle cell architecture and organization in *M. leidyi* is unique, exhibiting an aboral–oral orientation (longitudinal muscles, see Figure 7F in Martindale and Henry, 1999), aligned with the direction in which the epidermis is constructed. Therefore, I speculate that mesenchymal cells harbor the niches for epidermis differentiation or, perhaps, even for nervous system differentiation.

ECM rigidity, understood as textural patterns that arise within the embryo, constitutes another decisive and crucial mechanical cue in cell differentiation. During gastrulation, the presence of ECM textural gradients may influence cell trajectories (Loganathan et al. 2012; Hartman et al. 2017; Park et al. 2018; Zhu et al. 2020, 2023). On a spatial scale, qualitative differences in texture (rough versus smooth) can result in differential cellular outcomes. Under conditions of mechanical stress, mesenchymal stem cells commit to neural, muscular, or osteogenic lineages, depending directly on ECM elasticity and MyoII activity (Engler et al. 2006). In agreement with Engler's findings, oral micromeres in ctenophores can differentiate into neural cells (Jager et al. 2011), muscle cells (Martindale and Henry 1999), and mesenchymal cells of the mesoglea (Pang and Martindale 2008) after migration. In *Pleurobrachia pileus*, two types of nerve networks have been described: a loosely organized network distributed throughout the mesoglea and a more compact network with polygonal meshes located in the ectodermal epithelium (Jager et al. 2011). In vertebrates (Mongera et al. 2019; Sambasivan and Steventon 2021; Wymeersch et al. 2021) and ascidians (Hudson and Yasuo 2021), it is well established that certain posterior neural tissues share a common origin with the mesoderm, despite differences in mechanisms of specification (Hudson and Yasuo 2021). Based on these observations, it seems reasonable to propose that a neuromesodermal lineage has been conserved since the origin of metazoans. In vertebrates, the spatiotemporal localization of this lineage generally coincides with the co-expression of the transcription factors brachyury/TbxT and Sox2 (Aires et al. 2018). Brachyury has been characterized in ctenophores and cnidarians (Yamada et al. 2007, 2010; Pukhlyakova et al. 2018). Sox2 has been identified in ctenophores in regions of intense cell proliferation (Schnitzler et al. 2014; Moroz 2015) and in neurosensory epithelia (Jager et al. 2008).

Considering the current state of the art in cell differentiation, I wish to offer two reflections. The first is that, despite the apparent reductionism of our experimental approaches in embryology and the seeming impossibility of fully grasping the complexity of embryonic development as a whole, the most important question in developmental biology inevitably arises: How is embryonic robustness achieved? (Mao and Wickström 2024). The answer to this question lies in the field of physics. In my view, the mechanical memory embedded in the zygote (totipotency) may explain the robustness and coherence of tissue emergence and its mechanical interactions. Such knowledge will be crucial for translating advances in mechanobiology into practical tools for regenerative medicine (Cofre 2025). I will address the issue of totipotency later in this article. However, I urge caution regarding attempts to draw direct parallels between hydrogels of variable plasticity, gastruloids, and embryoids (Davies 2017; Shahbazi and Zernicka-Goetz 2018; van den Brink et al. 2020; Veenvliet et al. 2020; Wei et al. 2023; Lee et al. 2026; Hao et al. 2026) and the embryo in vivo. Such comparisons are akin to equating an intriguing laboratory experiment with a phenomenon shaped by evolutionary and phylogenetic construction. For this reason, I have left aside analyses based on these embryo-like systems, although I acknowledge their value in addressing specific therapeutic challenges. In this article, I focus instead on resolving broader questions related to totipotency, morphogenetic fields, and persistent mechanical memory rooted in animal phylogeny.

The second reflection concerns the true meaning of some concepts widely emphasized in the scientific literature, such as genetic programming (lineage-specific transcriptional programs) (Kalukula et al. 2025), molecular clocks (Cooke and Zeeman 1976; Palmeirim et al. 1997; Dubrulle and Pourquié 2002; Pourquié 2003; Andrade et al. 2007; Hubaud et al. 2017; Klepstad and Marcon 2023), and the so-called default state (Weinstein and Hemmati-Brivanlou 1997; Hemmati-Brivanlou and Melton 1997; Muñoz-Sanjuán and Brivanlou 2002). The following questions arise as we seek to highlight the role of mechanobiology and its future impact on the innovation of therapeutic strategies based on cell differentiation. What is the impact of using stem cells as platforms to study the dynamics of morphogenesis in vitro without considering relationships among cells within the integrated, global context of the embryo? What constitutes the self-organization of stem cells and their descendants? How do the extremely refined curvatures and shapes of different organs emerge during animal phylogeny? How are human forms consistently preserved and recreated across generations (related to robustness, as addressed in the first reflection)? Finally, the most important question in experimental biology: What is totipotency? In my view, all these inquiries can be addressed within the framework of persistent mechanical memory (as will be discussed later) and through an understanding of how physics directly influences biological phenomena.

The following section shows that molecular signals of nervous system induction are conserved in invertebrates (Gilbert 2000). This finding supports the discussion of the possibility that the so-called default state of neural differentiation may, in fact, represent a persistent mechanical memory that conserves and stores the mechanical records of animal morphogenesis (self-organization). In the same way, the so-called genetic programs and the self-organization of stem cells may be understood as consequences of tissue mechanics directly impacting gene expression (Tajik et al. 2016; Vining and Mooney 2017; Carley et al. 2021; Sun et al. 2023).

## 3. Instead of chemical signals, mechanical records serving as persistent mechanical memory: the physical foundations of nervous system differentiation

The formation of the nervous system represents one of the most enigmatic aspects of ctenophore embryology, and its possible independent origin has generated considerable controversy (Moroz et al. 2014; Marlow and Arendt 2014). The cellular basis of the default theory of neural induction are well established (Levine and Brivanlou 2007) and disregard the integration of mechanical information. Although there are proposals regarding the mechanical control of nervous system formation (Franze 2013), there remains a reluctance to speak explicitly and directly about mechanics in new hypotheses concerning the neural default model in vertebrates (Sagha 2025). In my opinion, incorporating elongation into the neural default model means integrating mechanical phenomena and anisotropic stress dependence into the very foundations of the theory (Bertet et al. 2004; Kumar et al. 2006; Rauzi et al. 2008; Fernandez-Gonzalez et al. 2009). At this juncture, when genetic and chemical aspects are considered predominant, it is natural to focus on identifying genes responsible for the process or, at the very least, to investigate whether these genes are related to nervous system formation in primitive animals such as ctenophores. Surprisingly, nearly all of the relevant components are already present during early embryonic development. Tolloid and TGF-β display overlapping expression patterns, consistent with preferential epidermal induction in the aboral region, identified by Martindale as the tentacular bulb (Pang et al. 2011). BMP homologs are likewise detected in the aboral region, where they are associated with the apical organ (Pang et al. 2011). The nervous system exhibits a distribution pattern typical of bilaterians, characterized by an aboral neurosensory complex (Jager et al. 2011). However, genes encoding BMP inhibitors such as *chordin* and *follistatin* are missing (Pang et al. 2011), as are other BMP and TGF-β inhibitor genes known to function in the endoderm region (Henry et al. 1996). When considering physical phenomena, what matters is not which structure performs the force or generates the tension, but rather which cellular component carries it out at any given moment in evolutionary history. From my point of view, the element that replaces chordin at the root of animal phylogeny is the ECM. However, this proposal implies a radical conceptual shift in which the geometry, textures, and autonomous forces of the ECM become fundamental to cellular differentiation. For this to occur, we must abandon the idea of morphogenesis driven solely by cells and instead focus on the dynamics and autonomy of the ECM (Loganathan et al. 2016; Nematbakhsh et al. 2020; Díaz-de-la-Loza and Stramer 2024).

The first step in reinterpreting the role of the ECM was breaking free from the trap of purely chemical interactions and their presumed direct effects on cellular differentiation. First and foremost, it was essential to identify all the interactions occurring around the ECM. Several ECM components have been shown to bind to BMP. Type IIA procollagen, which contains a cysteine-rich propeptide, binds to TGF-β1 and BMP-2 (Zhu et al. 1999). Collagen IV also binds to BMP-2 (Paralkar et al. 1992) and decapentaplegic (Dpp), the BMP-2/4 ortholog in *Drosophila melanogaster* (Wang et al. 2008). It then became important to recognize that two cysteine-rich domains alone are sufficient, rather than the entire chordin protein, for inhibiting BMP (Larrain et al. 2000). Moreover, cysteine-rich domains in other intracellular proteins have been identified as one of the major innovations of unicellular holozoans (Herman et al. 2018), configured as an important element in the formulation of a mechanical preneural model in an evolutionary context.

But the most significant breakthrough was achieved by establishing that proteins containing cysteine-rich domains, such as integrins, undergo conformational changes that expose cryptic sites not predicted in their initial structure; therefore, the dynamics through which these proteins interact appear to be highly important (Beglova et al. 2002). The idea emerged that the proteolytic processing of certain proteins implicated in embryonic development could reveal new biochemical activities associated with BMP inhibition (Yu et al. 2000). With this paradigm shift, the architecture of the ECM gained fundamental importance, and the basement membrane was understood not merely as a site for the deposition of protein fragments and growth factors, but as a site of biomechanical regulation (Kopf et al. 2014). In other words, BMP growth factor signaling and its impact on epidermal, osteogenic, and nervous system differentiation are only possible within a biomechanical context. The ECM was found to behave as an active agent in developmental biology, something that, for some developmental biologists, was extremely promising in terms of harnessing the driving force imparted by the embryonic environment (ECM) for future cell therapies (Bronner-Fraser 1982).

Donald Ingber's proposal of geometric control over cell life and death was a turning point in our understanding that cellular adhesion to micropatterned substrates is decisive for cell fate (Chen et al. 1997). Geometric foundations make sense in the fields of developmental biology and oncology (Kim et al. 2020), pointing to the idea that the real impact of morphogenesis is to reveal cell differentiation. Thus, stem cells appear to define their fate by taking advantage of nanotopography and integrin–matrix interactions (Kim et al. 2012; Dalby et al. 2014). Consistent with mechanical differentiation, chondrogenic progenitors or adipogenic precursor cells, which exhibit a spherical phenotype, fail to differentiate into bone cells (osteogenic fate) when exposed to local BMP gradients (Wang et al. 2012). These observations shed light on the existence of a synergistic mechanism between chemical (BMP-2) and mechanical signals (Tan et al. 2014; Wei et al. 2015, 2020). What was truly inspiring was the recognition of a molecular memory unit (Wei et al. 2020). Smad complexes remain bound to and active at target genes even after removal of BMP-2. In other words, the biomechanical impact can be measured at the level of gene transcription. There is already a consensus that the role of BMP as a morphogen, and its relationship with mechanotransduction, is embedded in the ECM and its remodeling, all within a surprising context in which chemical morphogens seem to have been replaced by mechanical morphogens (Das et al. 2019).

In the presence of collagen IV, BMP diffusion coefficients decrease from about 87 to 0.10 $\mu m^2\ s^{-1}$ (Sedlmeier and Sleeman 2017), significantly limiting the range predicted by chemical models based on diffusion gradients (Turing 1952; Wolpert 1969, 1971; Rogers and Schier 2011). Today, there are genuine doubts as to whether BMP ligands act as true morphogens. Dpp is one of the most classically characterized morphogens in developmental biology. Yet, diffusion analyses in wing imaginal discs of *Drosophila* have shown only short-range effects (Kicheva and González-Gaitán 2008; Ramel and Hill 2012), consistent with findings in vertebrates such as *Xenopus* and zebrafish (Jones et al. 1996; Nikaido et al. 1999). Also, recent studies have concluded that transcytosis is not involved in Dpp transport in the *Drosophila* wing (Schwank et al. 2011), confirming its short-range action. Therefore, here we have a morphogen (Dpp) that does not freely diffuse but that, to achieve long-distance effects, could act synergistically with mechanical mechanisms in morphogenesis (Nematbakhsh et al. 2020; Geresu et al. 2025) and growth (Aegerter-Wilmsen et al. 2012; Baena-Lopez et al. 2012) in the imaginal disc of *Drosophila*.

The importance of mechanics in cell differentiation was consolidated by studies showing that, in response to compressive force, integrins bind synergistically to the BMP receptor and form a mechanoreceptor complex that mediates Smad1/5/8 phosphorylation (Zhou et al. 2013). This finding revealed that the initial component of the mechanism is entirely mechanical and involves oscillatory shear stress. Finally, the BMP-1 receptor interacts with the integrin β1 subunit to induce Smad1/5 signaling in response to increased substrate stiffness (Guo et al. 2016), paving the way for recognition of the importance of textural gradients and anisotropic textures in cell differentiation (Swift et al. 2013; Zouani et al. 2013; Shekaran et al. 2014; Chaudhuri et al. 2020; Sthanam et al. 2022; Alisafaei et al. 2025) and in cancer (Mott and Werb 2004; Park et al. 2016; Ahmadzadeh et al. 2017; Glentis et al. 2017; Yan et al. 2017; Dai et al. 2020b; Wang et al. 2020; Zhang et al. 2020; Naylor et al. 2022). Based on these fundamentals, I consider the ECM to represent a major revolution in animal organization (Cofre 2023), not only for its possible ability to substitute chordin in ctenophores or for its mechanotransducing power, but mainly for its now well-established capacity to generate persistent mechanical memory (Guilak et al. 2009; Balestrini et al. 2012; Yang et al. 2014; Heo et al. 2016; Li et al. 2017; Killaars et al. 2019, 2020). According to my hypothesis, persistent mechanical memory is what makes it possible to reproduce an embryo in the next generation. Furthermore, it is responsible for triggering the emergence of totipotency (Figure 1). Next, based on all the scientific arguments described above, I will show how a mechanical model of neural default could arise in the first animal embryo.

During embryogenesis, ECM networks display highly dynamic material properties that accommodate the large-scale deformations and forces shaping the embryo. In epithelial cancer cell models, tension waves are generated and propagate over long distances (Serra-Picamal et al. 2012) in the direction of migration, functioning as "morphogens" (Morita et al. 2017; Das et al. 2019; Agarwal and Zaidel-Bar 2021). In ctenophores, the nervous system is predominantly located in the aboral region, reflecting the effects of early patterning (Jager et al. 2011); however, this organization has not always been so. *Ctenorhabdotus* *campanelliformis* possessed longitudinal axons connecting the apical organ and ciliated sulci (aboral) to a circumoral (oral) nerve ring (Parry et al. 2021). The ECM microenvironment also changes markedly over time and space during morphogenesis, generating fluctuations in textural properties (Loganathan et al. 2012; Hartman et al. 2017; Park et al. 2018; Zhu et al. 2020, 2023). These textural gradients arise from the physical forces involved in the self-organization of the first embryo. Based on these considerations, I speculate that the ECM sequesters BMP according to textural patterns, creating regions where limited diffusion of BMP induces epidermis formation (Wilson and Hemmati-Brivanlou 1995; Zimmerman et al. 1996) and regions without BMP (retained within the ECM) (Figure 2A and B) and thereby allowing default formation of the nervous system (Levine and Brivanlou 2007). Here it is important to emphasize that long-range diffusion is not required; only BMP secretion is necessary to generate a local textural gradient near the secretory cell during ctenophore epiboly. In my proposal, this textural gradient, acting in synergy with the stress waves produced by epiboly (from oral to aboral), would allow the induction of cell morphogenesis and differentiation. In line with this idea, it has been proposed that the polygonal network pattern of the nervous system would follow the textural organization of the ECM in the absence of BMP-4 diffusion (Jager et al. 2011). This same textural pattern might have been important for the induction of a neuromesodermal lineage during gastrulation.

To develop the preneural default model, we must imagine the first embryo and the earliest instance of cell differentiation in the metazoan context. In this framework, the preneural model is grounded in mechanics and represents the earliest manifestation of durotaxis and topotaxis, phenomena now well described as biophysical regulators of migration and morphogenesis (Engler et al. 2006; Sunyer et al. 2016; Park et al. 2018). It is equally important to recognize that epiboly entails a redistribution of viscoelastic forces acting on macromeres, together with an opposing surface tension that pulls cells toward the oral pole, consistent with models derived from cancer research (Pajic-Lijakovic and Milivojevic 2020, 2022). As epiboly progresses, this tension increases because the number of cells expands, the cell layer enlarges (Martindale and Henry 1999), and the ECM is progressively incorporated (Latimer and Jessen 2010). These changes result in greater force exerted on the substrate and heightened tension between cells in the oral region (Figure 1A). At the same time, one can imagine a stronger viscoelastic resistive force acting on macromeres at the aboral pole. The model assumes that cells undergoing epiboly generate intense tensile forces and are capable of establishing the ECM textural gradient by themselves (Figure 2A and B). If the thin ECM layer were removed after epiboly, the resulting profile would resemble that shown in Figure 2C. Such a profile reflects the tensile forces applied to cell surfaces (Figure 1A), which shape larger collagen deposits at sites of cell–cell interactions, thereby increasing ECM rigidity (stiffening) relative to softer surrounding regions (Figure 2C).

Following epiboly, ectodermal cells proliferate while becoming exposed to the established textural gradient, which contributes to the differential formation of the epidermis and nervous system (Wilson and Hemmati-Brivanlou 1995). Stress waves generated during epiboly propagate as morphogens (Serra-Picamal et al. 2012). In ctenophores, this mechanical polarity is partly reflected in the restricted expression of BMP homologs to the aboral region (Pang et al. 2011). Within the ectoderm, BMP secretion (Figure 2B) generates the profile illustrated in Figure 2D: regions with increased stiffness, enriched collagen content, and higher densities of cysteine-rich domains that sequester BMP, resulting in reduced local BMP concentration. Cells entering these regions would therefore initiate epidermal differentiation. Conversely, in regions without free BMP, neural differentiation, by default, would be induced, owing to the absence of BMP receptor activation (Figure 2E). The polygonal configuration shown in Figure 2F corresponds to the organization of neural cells.

This polygonal ectodermal network matches the nerve net architecture described in ctenophores, specifically the epithelial polygonal nerve network (Jager et al. 2011). Surprisingly, a mesogleal nerve network has also been reported in ctenophores (Jager et al. 2011), a finding consistent with the proposed model. As the thin ECM film and its textural gradient interact with two cell populations, they would also influence cells within the mesodermal layer, thereby promoting a neuromesodermal lineage. For readers familiar with textural gradients and mesenchymal differentiation, it is worth noting that the pattern proposed in Figure 2 aligns with parameters known to favor osteogenic differentiation (Engler et al. 2006; Huebsch et al. 2010; Tse and Engler 2011; Shih et al. 2011). The essential point is that cells, through epiboly, generate a textural gradient that begins to establish durotaxis and topotaxis. This understanding represents a significant advance in the mechanical challenges of embryogenesis as physical principles operating throughout animal phylogeny.

Returning to my hypothesis, macromeres subjected to tensile forces during gastrulation would differentiate into the endodermal layer and into mesodermal components of the mesoglea, including a neural network. Here, I would like to emphasize that embryogenesis itself constructs tissue architecture and organization. If the embryonic process is not taken into account, totipotency will continue to be defined merely as the capacity to generate distinct cell types, rather than as the outcome of the embryo's ability to build tissue architecture and organization during morphogenesis. From this perspective, epiboly, invagination, and other morphogenetic movements are directly involved in revealing totipotency.

## 4. Totipotency revealed through morphogenesis

Defining totipotency solely as a problem of differentiation, without taking morphogenesis into account, appears to stem from the limitations of experimental approaches.

Studying cell differentiation often involves disrupting the embryo and its organization, thereby losing the physical and mechanical context that shapes the process (Discher et al. 2009). A clear example of the pitfalls of this approach is the widespread belief that pluripotent embryonic stem cells provide a unique window into mammalian development and differentiation in a Petri dish (Rossant 2008), or that isolated embryonic stem cells can serve as the basis for new models of mammalian development (Murry and Keller 2008; Young 2011). Whereas the ultimate goal is to control the differentiation potency of embryonic stem cells and steer their development along specific pathways, this strategy overlooks crucial information encoded in the embryonic cells themselves, such as their niches, interactions with the ECM, and morphogenetic movements, which may hold the key to understanding the potency problem.

Another frequent misconception is to attribute the morphogenetic construction of the embryo solely to genes (Gehring 1996). In vertebrates, gastrulation genes are often treated as central components of genetic strategies in developmental biology. Yet, when gastrulation occurs in the absence of these genes, driven instead by physical forces, which are rarely recognized as active agents of morphogenesis (Brunet et al. 2013; Mitrossilis et al. 2017; Pukhlyakova et al. 2018; Bailles et al. 2019), our technical and theoretical limitations become apparent. Genes are also commonly credited with establishing differentiation potential, as exemplified by induced pluripotent stem cells (iPSCs) (Takahashi et al. 2007). Although the contributions of iPSC research to basic and applied science are undeniable, the limitations of these models are often overlooked, particularly with respect to differentiation protocols that require refining (Ratajczak et al. 2016; Antonov and Novosadova 2021). The links between reprogrammed embryonic stem cells and cancer are also poorly understood (Hwang et al. 2009; Lin et al. 2014; Du et al. 2020; Seno et al. 2022). Nevertheless, it is known that *Oct3/4* (de Jong and Looijenga 2006), *Sox2* (Dey et al. 2022), *Klf4* (Qi et al. 2019), and *Myc* are linked to cancer.

Accordingly, iPSCs injected intraperitoneally form teratomas, displaying pluripotent features (NANOG expression) consistent with their embryonic identity (Abad et al. 2013).

Despite our experimental limitations, understanding how cells retain a memory of their mechanical history, whether lasting minutes, days, or weeks, has emerged as a crucial aspect of embryology. The physical and mechanical experiences that guide the successful construction of the first embryo may be stably and persistently impregnated in stem cells or germ cells as mechanical memory, potentially allowing the process to be reconstructed in subsequent generations. This insight opens the possibility of integrating physical principles directly into the concept of totipotency. Current research continues to investigate the epigenetic mechanisms responsible for the formation and maintenance of this mechanical memory (Heo et al. 2015; Scott et al. 2023; Cambria et al. 2024; Lee and Holle 2024). It is well established that epigenetic processes contribute to the storage of mechanical memory in cells, supporting the consensus that mechanical forces can modify chromatin accessibility and influence transcriptional activity (Rao et al. 2014; Le et al. 2016; Kim et al. 2019).

It is now recognized that long-term mechanical memory can be encoded through persistent epigenetic modifications (Balestrini et al. 2012), including histone methylation and acetylation, which are regulated by methylases, demethylases, histone acetyltransferases, and histone deacetylases, as well as DNA methylation and non-coding RNAs (Cambria et al. 2024). The most compelling evidence of these persistent epigenetic mechanisms comes from embryology. Polycomb group (PcG) proteins, key epigenetic modifiers, are central to genetic repression and play critical roles in regulating developmental genes across multiple cell types and tissue contexts, including embryonic and adult stem cells. Furthermore, these proteins are essential for cell fate transitions during embryogenesis (Sparmann and van Loewen 2006) and have specific roles in embryonic development, in the functioning of pluripotent stem cells, and in the reprogramming of somatic cells to a pluripotent-like state (Aloia et al. 2013).

Histone acetylation is generally linked to chromatin opening and gene activation (Bannister and Kouzarides 2011). An important exception to this rule is H3K27me3, which promotes gene repression (Bracken et al. 2007; Piunti and Pasini 2011). PRC2-mediated trimethylation of H3K27 can enhance the binding affinity of local chromatin for Polycomb repressive complex 1 (PRC1) (Min et al. 2003). When PRC1 binds to H3K27me3 marks, heterochromatin is formed, restricting chromatin accessibility to transcriptional complexes and inhibiting subsequent transcription via chromatin compaction (Grimaud et al. 2006; Schuettengruber et al. 2017). In vitro, exposure of multipotent epidermal stem/progenitor cells and mesenchymal stem cells to cyclic mechanical tension induced chromatin compaction and transcriptional repression dependent on H3K27me3 (Le et al. 2016; Heo et al. 2016), phenomena that may underlie the formation of persistent mechanical memory. Moreover, increasing mechanical tension resulted in the amplification of the extent and persistence of chromatin condensation in mesenchymal stem cells, revealing the dose-dependent nature of these repressive effects (Heo et al. 2016). Altogether, these findings suggest that embryonic stem cells possess mechanisms capable of establishing and maintaining long-term mechanical memory, a feature indispensable for the emergence of animal life.

It was only possible for animal life to arise once all essential mechanical processes were fully operational in the first embryo. The earliest zygote needed to transmit the physical consequences of intense reorganization of the actin cytoskeleton to the nucleus, thereby organizing chromatin. Cells undergoing epiboly and those adhering to the basal lamina had to undergo the same process to withstand the physical stresses of morphogenesis. Within a short time, a population of multipotent stem cells acquired the ability to absorb and endure mechanical impacts throughout their ontogeny. This capability endowed stem cells with a biomechanical history, stored as mechanical memory within their nuclei and later used to reproduce the process in subsequent generations (Figure 1B).

Basal animals retain germ cells within the pool of multipotent stem cells for an extended period (Juliano et al. 2010; Fierro-Constaín et al. 2017; Edgar et al. 2021), delaying their segregation in a process reminiscent of cancer stem cell behavior (Cofre 2025). This delayed separation was both expected and crucial for the emergence of animal life. Multipotent stem cells in the first embryo required mechanical and physical stimuli from embryogenesis to complete germline segregation. In other words, germ cells could only form after being fully imprinted by their surrounding environment and the physical processes occurring within the first embryo (Cofre and Saalfeld 2023). Through this process, the embryo encoded the genomic records of its construction by generating mechanical memory (Engler et al. 2006; Heo et al. 2015; Dai et al. 2020) within multipotent stem cells. This principle applies to the first embryo, which created a topological map of physical forces within its genetic material (Uhler and Shivashankar 2017b, a, 2018; Tsai and Crocker 2022).

Building on the concept of mechanical memory, Denis Duboule's studies with HoxD genes in embryonic stem cells showed that non-transcribed genes are marked by H3K27me3 tags, whereas active genes carry large domains of H3K4me3 tags (Soshnikova and Duboule 2009). This relationship suggests that bimodal 3D dynamics may have a mechanical basis, considering the H3K27me3-dependent sensitivity to mechanical tension (Heo et al. 2016). It is also notable that the multidirectional interactions of topologically associating domains (TADs) were retained after cohesin depletion (Bintu et al. 2018), further hinting at their mechanical underpinnings. Cohesins themselves are force-sensitive (Kim et al. 2019; Pobegalov et al. 2023) and play critical roles in the pairing of sister chromatids during mitosis (Strunnikov et al. 1993; Guacci et al. 1997; Michaelis et al. 1997; Srinivasan et al. 2018), as well as in loop extrusion (Richeldi et al. 2024; Golovand Gavrilov 2024) after cytokinesis. Chromatin features, including higher-order organization (e.g., TADs and Polycomb domains) and frequent cooperative interactions, have drawn significant attention as emergent properties (Bintu et al.

2018; Almassalha et al. 2025), a form of developmental memory (Noordermereer et al. 2014), or even as a reinforcement learning system (Almassalha et al. 2025). Taken together, these observations indicate that the structural organization of chromatin is imbued with long-term persistence, and that chromatin dynamics provide a foundational framework for beginning to explain totipotency.

## 5. Concluding remarks and perspectives

At present, from the perspective of the history of science, it is essential to recognize and test the mechanical component of totipotency, as this could have profound implications for both cancer therapy and our understanding of animal evolution from a physics standpoint. Today, it is possible to quantitatively test the hypothesis that totipotency involves mechanical processes in animal zygotes using transcriptomics (considering effects on gene transcription within the context of TADs), an approach congruent with the philosophy of physics (Carnap 2012). Quantitative measurements of mechanical forces in the zygote and early embryonic stages may allow the investigation of emergent properties, such as the formation of migratory epithelia during epiboly in invertebrates, with the potential to reduce these phenomena to physicochemical principles. From a physics standpoint, it is crucial to identify a minimal set of non-linear interactions that are both necessary and sufficient for the emergence of specific embryonic properties. By applying this strategy, we can integrate physics into organismal biology from a cancer perspective (Cofre 2025), using in vivo models without decomposing the system. Recognizing the inherent limitations of functional genomics (Medina 2005; Boogerd et al. 2007) may bring us closer to developing fully testable and falsifiable theories of embryogenesis, thereby establishing embryology as a solid framework for exploring physical principles.

A physical perspective on cell differentiation invites a paradigm shift. Evolutionary biology and the study of biological diversity (Gilbert 2003) have, incomprehensibly, distanced themselves from physics. Biologists have largely ignored, or even renounced, the task of understanding how animal forms generate themselves (Bolker 1995; Ball 2015). Even the bold attempt to acknowledge that understanding the creation of any animal form would be impossible without knowledge of physics has often been overlooked (Keller 2012; Green and Batterman 2017). I foresee several consequences of adopting a physical approach to embryology (developmental physics) in modern medicine. If physicists and biophysicists engage with cell differentiation, embryology, and cancer, the focus in breast cancer therapy may shift away from genes (Puget et al. 2002) toward textural gradients (Dai et al. 2018, 2020b; Ren et al. 2019). In my proposal, the embryo serves as a biological inspiration for designing matrices aligned with a dynamic, autonomous ECM capable of generating forces (Loganathan et al. 2016). In this view, the biomechanical structure of morphogenesis is what drives cell differentiation. Another potential medical application would be studying the physical phenomena associated with the constant growth and retraction of the breast during a woman's menstrual cycle, which could inform both prevention and therapy for breast cancer. Meanwhile, the study of textural gradients in embryonic models in vivo is already possible (Zhu et al. 2025). Mapping tissue textures, forces, and tension in embryos is part of our current reality, and only in the future will we fully unravel the therapeutic potential of embryonic mechanobiology.

**Acknowledgments**

I thank Dr. Jose Bastos, pathologist and oncologist, for his ongoing contributions to my cancer research projects. I apologize to any researcher who was not cited in the article, given the necessity of prioritizing references in its construction.

**Conflict of interest statement**

This research was conducted in the absence of any commercial or financial relationships that could be construed as potential conflicts of interest.

Figures

Figure 1

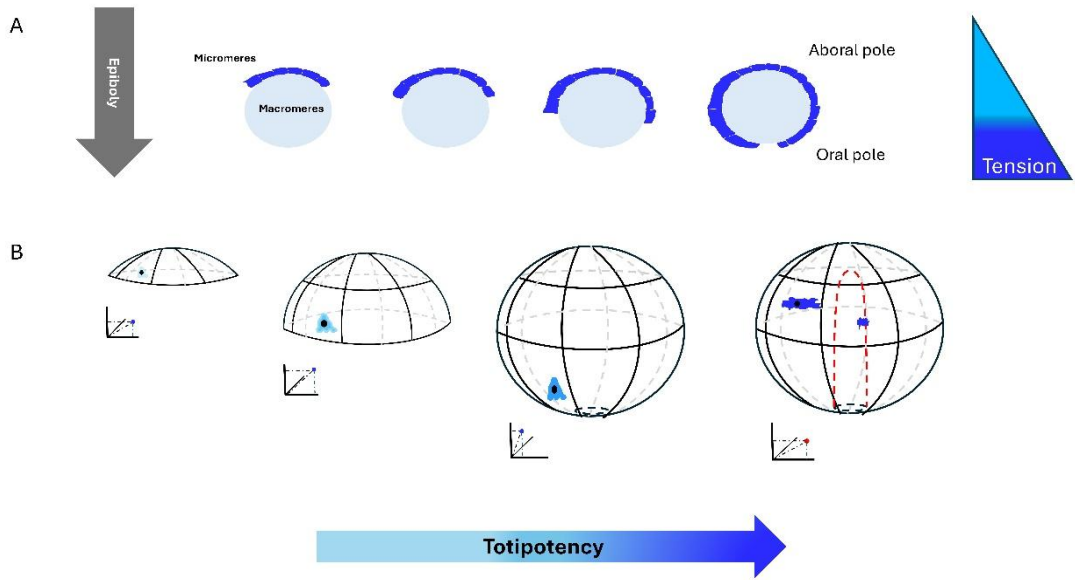

Figure 2

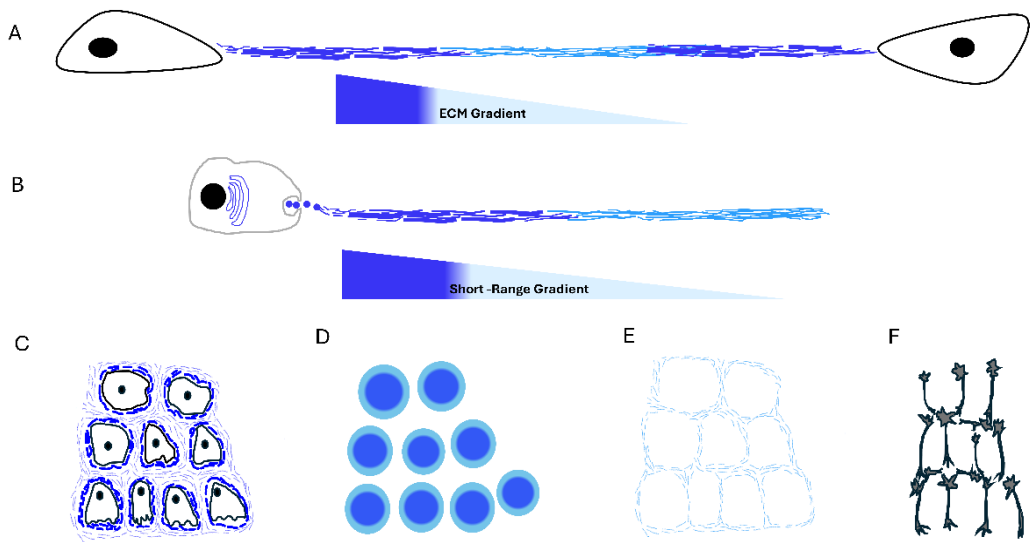

Legends

**Figure 1.** Totipotency was acquired through mechanical impact during morphogenesis, thereby creating a persistent mechanical memory during the construction of the first animal embryo. The figure represents the hypothetical first embryogenesis (that is, before the first animal had been formed). It is proposed that the first embryogenesis arose from protist cells, from which animal phylogeny emerged. (A) General representation of epiboly in ctenophores, basal animals placed at the root of animal phylogeny. Cells performing epiboly do so in direct contact with macromeres. Cell movement occurs in the aboral-to-oral direction, with an increase in mechanical tension in the expanding epithelium. (B) As embryogenesis proceeds, stem cells (blue dots within the geometric network) record the mechanical history of the embryo as a persistent mechanical memory, which will allow the process to be reproduced in the next generation. The next generation is the first zygote produced by fertilization of animal germ cells, themselves generated by the first hermaphroditic embryo. The first animal zygote would express the persistent mechanical memory constructed during the first embryogenesis, as depicted here. For this to occur, the transition from somatic to germline lineage must happen late, after the full mechanical record of the first embryogenesis has been established. The gradual development of totipotency is represented by an increasing intensity of blue in the figure.

**Figure 2.** Biophysical model describing the default preneural induction of the nervous system. The model is based on the proposal that cells undergoing epiboly generate intense tension forces and are capable of creating the textural gradient of the extracellular matrix (ECM) by themselves. The ECM acts synergistically with chemical signals (bone morphogenetic proteins, BMPs) secreted by cells. (A) Stiffness profile of the ECM of cells that underwent epiboly, produced by the stresses experienced by cells. (B) A gradient of BMPs, which trigger the formation of the epidermis and the ventral mesoderm, is produced and strongly accentuated by the ECM. Collagen IV sequesters BMPs and helps create a short-range BMP gradient. (C) It is possible to observe a region with low BMP concentration (dark blue) near cells at the end of epiboly and a region without BMP concentration (light blue) farther away. (D) The only region

with available BMPs (limited free BMPs) surrounds the cell, being in close contact with it. This organization allows for the formation of discrete spots of epidermal induction without connections between them. (E) An interconnected ECM network without BMPs (low stiffness) is observed, which will be used for neural differentiation. (F) Polygonal profile generated by textural differences in the ECM. The nervous system differentiates in regions lacking BMPs. This profile corresponds to the epidermal polygonal nerve network of ctenophores (Jager et al. 2011).